\documentclass[a4paper,usenatbib]{mnras}

\usepackage{graphicx}
\usepackage{amsmath}
\usepackage{amssymb}

\title[WISE Planet Nine Search]{A 3$\pi$ Search for Planet Nine at 3.4$\mu$m with WISE and NEOWISE}

\author[Meisner et al.]{
A.~M. Meisner,$^{1,2}$\thanks{ameisner@lbl.gov}
B.~C. Bromley,$^{3}$
S.~J. Kenyon,$^{4}$
T.~E. Anderson$^{3}$
\\
$^{1}$Berkeley Center for Cosmological Physics, Berkeley, CA 94720, USA \\
$^{2}$Lawrence Berkeley National Laboratory, Berkeley, CA, 94720, USA \\
$^{3}$Department of Physics \& Astronomy, University of Utah, Salt Lake City, UT, 84112, USA \\
$^{4}$Smithsonian Astrophysical Observatory, 60 Garden Street, Cambridge, MA 02138, USA \\
}

\pubyear{2017}

\begin{document}
\label{firstpage}
\pagerange{\pageref{firstpage}--\pageref{lastpage}}
\maketitle

\begin{abstract}
The recent `Planet Nine' hypothesis has led to many observational and archival searches for this giant planet proposed to orbit the Sun at hundreds of astronomical units. While trans-Neptunian object searches are typically conducted in the optical, models suggest Planet Nine could be self-luminous and 
potentially bright enough at $\sim$3--5$\mu$m to be detected by the Wide-field Infrared Survey Explorer (WISE). We have previously demonstrated a Planet Nine search
methodology based on time-resolved WISE coadds, allowing us to detect moving objects much fainter than would be possible using single-frame extractions. In the present work, we extend our 3.4$\mu$m (W1) search to cover more than three quarters of the sky and incorporate four years of WISE observations spanning a seven year time period. This represents the deepest and widest-area WISE search for Planet Nine to date. We characterize the spatial variation of our survey's sensitivity and rule out the presence of Planet Nine in the parameter space searched at W1 $< 16.7$ in high Galactic latitude regions (90\% completeness).


\end{abstract}

\begin{keywords}
surveys: trans-Neptunian objects --- methods: data analysis --- techniques: image processing
\end{keywords}

\section{Introduction}
\label{sec:intro}

Over the past fifteen years, observations have emerged which challenge the conventional understanding of our solar system's outer regions. Chief among these
are the discoveries of Sedna \citep{brown04} and 2012 VP$_{113}$ \citep{trujillo14}, trans-Neptunian objects (TNOs) with orbits that cannot be explained by perturbations from the solar system's known planets \citep{gladman2002, morbidelli2004, brown04}. These two objects are often grouped together with other high perihelion ($q > 30$ AU), large semi-major axis ($a > 250$ AU) bodies into a class of objects referred to as extreme TNOs (ETNOs). Thirteen such ETNOs are now known \citep{brown04, trujillo14, sheppard16, bannister17}. These objects have been found to cluster in argument of perihelion \citep{trujillo14} or longitude of perihelion \citep{batygin16, brown17}. To explain the clustering in argument
of perihelion, \cite{trujillo14} proposed
the existence of a super-Earth mass perturber at $\sim$250 AU. In a related but different analysis, \cite{batygin16} suggested that an unseen giant planet (`Planet Nine')
of mass 5--20M$_{\oplus}$ can simultaneously explain many enigmatic features of the solar system. These include the longitude of perihelion clustering among ETNOs, the existence of Sedna and 2012 VP$_{113}$, the population of high-inclination TNOs \citep{batygin16b}, and the solar obliquity \citep{obliquity, gomes17, lai16}. Criticisms of this Planet Nine
narrative have also been put forth. For instance, the  ETNO clustering in longitude of perihelion has been argued to result from observational bias \citep{shankman17}, and the presence of such a massive unseen planet has been claimed inconsistent with Cassini ranging measurements \citep{folkner}. \cite{lawler2017} and \cite{nesvorny2017} did not find evidence that ETNO clustering in $\omega$ or perihelion longitude would result from the existence of a body like Planet Nine. \cite{nesvorny2017} also suggested that Planet Nine would create a distribution of ecliptic comet inclinations in tension with that observed.


Nevertheless, the Planet Nine hypothesis yields concrete predictions \citep{brown16} for the proposed planet's orbital parameters (380 $\lesssim$ $a$/AU $\lesssim$ 980, $i \sim 30^{\circ}$, 0.3 $\lesssim e \lesssim$ 0.75), facilitating searches which can directly observe or else place limits on its presence. Planet Nine is expected to
be faint in the optical, but not prohibitively so \citep[22 $<$ $V$ $<$ 25;][]{brown16}. Studies have also suggested the possibility of detecting Planet Nine at a range of other wavelengths \citep{cowan16, fortney16, linder16}, depending on its detailed physical properties. Owing in part to these encouraging forecasts, many archival and/or observational searches for Planet Nine are already underway  \citep{sheppard16, panstarrs, fortney16, kuchner17, p9w1_cetus}.


In this work, we focus on searching for Planet Nine in the infrared. Evolutionary models of Planet Nine spanning 5-60M$_{\oplus}$ predict that it could be self-luminous
\citep{linder16, fortney16}.  In particular, \cite{fortney16} suggested that Planet Nine may be much brighter than blackbody expectations
at 3--5$\mu$m, based on a detailed analysis of model atmospheres. \cite{fortney16} found that Planet Nine may be detectable by the Wide-field Infrared Survey Explorer (WISE)
in its bluest channel \citep[W1;][]{wright10}, although the models span an enormous range of luminosities at mid-infrared wavelengths.

The W1-brightest model from \cite{fortney16} still remains fainter
than the single-exposure detection limit, but would be readily detectable in coadds (at a fiducial distance of Planet Nine of $d_9 \sim$ 600 AU). In \cite{p9w1_cetus}, hereafter M17, we developed a methodology to convert the WISE W1 imaging 
into a serendipitous search for Planet Nine using a custom set of time-resolved coadd images. We applied this procedure to a $\sim$2,000 square degree region which, at the time, was
thought to represent a likely Planet Nine location \citep{holman16}. Here, we extend our W1 search to cover more than three quarters of the sky and include an additional year of WISE observations. Although several predictions have been made for Planet Nine's most likely current sky location \citep{brown16, holman16, dlfm16, millholland}, our present strategy is to uniformly search the widest area possible. The very high
and homogeneous quality of the space-based imaging delivered by WISE is conducive to deriving detailed constraints on Planet Nine's apparent mid-infrared brightness in the event of a non-detection.



In $\S$\ref{sec:data} we describe the archival WISE data used in the course of this search. In $\S$\ref{sec:recap} we review the Planet Nine search methodology presented in M17. 
In $\S$\ref{sec:footprint} we provide details of the footprint searched during the course of this work. In $\S$\ref{sec:orbfit} we explain our orbit linking procedure. In $\S$\ref{sec:completeness}
we discuss the sensitivity of our search. In $\S$\ref{sec:results} we describe the results of our search. We conclude in $\S$\ref{sec:conclusion}.

\section{Data}
\label{sec:data}

We begin by very briefly reviewing key aspects of the WISE mission and its timeline of observations. WISE is a 0.4 meter telescope onboard a satellite in low-Earth orbit \citep{wright10}. The satellite was launched in late 2009. WISE has surveyed the entire sky in four broad bandpasses, centered at 3.4$\mu$m (W1), 4.6$\mu$m (W2), 12$\mu$m (W3) and 22$\mu$m (W4). WISE performed a seven month full-sky survey in all four of its channels starting in 2010 January. In 2010 August and September, WISE surveyed the sky in its three bluest bands only. It then carried out an asteroid hunting mission called NEOWISE until 2011 February, operating only in the W1 and W2 channels \citep{neowise}. It then entered a hibernation period of nearly three years, and recommenced survey operations in 2013 December. The post-hibernation mission phase has been dubbed NEOWISE-Reactivation (NEOWISER). WISE remains operational in W1 and W2, and has already completed roughly eight full sky passes throughout its first seven years in orbit.

NEOWISER publicly releases its data in chunks that each span one year of observations. These data releases consist of single-exposure images and source extractions, and are published on a roughly annual basis. Thus far three such data releases have been published, in 2015 March, 2016 March and 2017 June.

A major difference between the present work and the initial search presented in M17 is our inclusion of an additional year of NEOWISER exposures obtained
from the third-year NEOWISER data release, which became publicly available in 2017 June. In this work we employ all publicly available W1 Level 1b (L1b) single-exposure images
that fall within our search footprint described in $\S$\ref{sec:footprint}. Our search footprint is defined primarily by cuts on ecliptic latitude (denoted by $\beta$, $\S$\ref{sec:beta}) and background source density ($\S$\ref{sec:plane}). In particular, we restrict our analysis to $|\beta|$ $<$ 55$^{\circ}$. The L1b acquisition dates range from 7 January 2010 to 13 December 2016 (UTC). In other words, we use all pre-hibernation W1 L1b frames, plus those frames included in the first \textit{three} annual NEOWISER data releases. This amounts to $\sim$49 months of WISE observations. While the input data spans a roughly seven year time period, the time baseline at a given sky location is typically 6.5 years.

We have downloaded\footnote{\url{http://irsa.ipac.caltech.edu/ibe/data/wise}} local copies of all publicly available W1 L1b sky images (\verb|-int-| files) along with their corresponding uncertainty maps (\verb|-unc-| files) and bitmasks (\verb|-msk-|
files). We refer to one such group of three W1 files resulting from a single WISE pointing as a `frameset'. The coadds presented in this work make use of $\sim$6 million such framesets, for a total of $\sim$40 TB and $\sim$18 terapixels of single-exposure inputs.







\section{Overview of Search Methodology}
\label{sec:recap}


Our search strategy is modeled after \cite{brown15}, although our analysis begins at the pixel level rather than the catalog level. \cite{brown15} developed a method
for linking transients with arbitrary time sampling into Keplerian orbits for solar system objects beyond 25 AU. We employ a similar orbit linking approach which we describe fully in
$\S$\ref{sec:orbfit}. The \cite{brown15} analysis begins with large catalogs of transient candidates provided by the Catalina Real-time Transient Survey \citep[CRTS;][]{crts}. The starting point for our analysis is a set of $\sim$6 million single-exposure WISE/NEOWISE images. Therefore, much of our analysis is devoted to image processing steps which allow us to 
obtain a clean list of transients from the available WISE images. In brief, our method stacks W1 exposures into a set of epochal coadds, extracts transients via difference imaging, and
finally attempts to link these transients into Keplerian orbits beyond 250 AU. The following subsections briefly review each step of our Planet Nine search procedure. Full details
are provided in M17.

\subsection{Tiling}

Our search region ($\S$\ref{sec:footprint}) is composed of a series of `tiles', with each tile corresponding to a unique $\sim$2.5 square degree astrometric footprint. These are the footprints onto which
we reproject L1b exposures during coaddition. We adopt tile footprints matching those
of the WISE team's Atlas stacks \citep{cutri11}. The Atlas tiling covers the entire sky with a set of 18,240 astrometric footprints. Each Atlas tile footprint is 1.56$^{\circ}$ on a side, and is identified by a \verb|coadd_id| value. The \verb|coadd_id| is a string encoding the tile's central (RA, Dec) coordinates. For instance, the Atlas tile centered at (RA, Dec) = (324.8$^{\circ}$, $-$19.7$^{\circ}$) has \verb|coadd_id| = `3248m197'. The Atlas tile centers trace out a series of isodeclination rings. 
For our time-resolved coadds, we adopt a pixel scale matching that of the L1b images, 2.75$''$/pixel. Thus, each of our tiles has dimensions of 2048 pixels $\times$ 2048 pixels.
In our terminology, `tiles' are distinct from `coadds': there are exactly 18,240 unique tile footprints, whereas we have typically constructed 8--9 epochal coadds for each such tile footprint within our search region (see Table \ref{tab:n_epoch}). Note that although the 18,240 astrometric footprints of the Atlas tiles are unique, neighboring tiles overlap at their boundaries. The amount of overlap is $\sim$3$'$ near the celestial equator.

\begin{table}
        \centering
        \caption{Fraction of pixels in our search footprint region as a function of number of
coadd epochs.}
        \label{tab:n_epoch}
        \begin{tabular}{cc}
                \hline
                $n_{epochs}$ & $p(n_{epochs})$  \\
                 \hline
                0 & $3.08 \times 10^{-6}$ \\
                1 &  $1.13 \times 10^{-6}$ \\
                2 & $2.90 \times 10^{-6}$ \\
                3 & $3.98 \times 10^{-6}$ \\
                4 & $6.30 \times 10^{-6}$ \\
                5 & $7.98 \times 10^{-6}$ \\
                6 & $8.13 \times 10^{-5}$ \\
                7 & $5.35 \times 10^{-2}$ \\
                8 & $3.08 \times 10^{-1}$ \\
                9 & $3.85 \times 10^{-1}$ \\
                10 & $2.21 \times 10^{-1}$ \\
                11 & $3.23 \times 10^{-2}$ \\
                12 & $4.62 \times 10^{-6}$ \\
                \hline
        \end{tabular}
\end{table}



\subsection{Time Slicing}
\label{sec:slice}
We seek to extract transients from time-resolved coadds of the W1 imaging rather than from single exposures themselves. This choice has two main advantages: the
coadds are $\sim$1.3 magnitudes deeper than single frames, and are also much cleaner because fast transients such as asteroids, cosmic rays, and satellite streaks are eliminated. 

The WISE survey strategy is such that a given sky region is `visited'  for roughly a day-long period once every six months while WISE is operational. 
We segment W1 exposures into such visits, stacking together those frames grouped within the same visit. The typical timespan
of such a `coadd epoch' at a particular sky location is roughly\footnote{This formula is not correct at very high ecliptic latitude, but is applicable over the $|\beta| < 55^{\circ}$ region analyzed in this work.} (1 day)/cos($\beta$). We implement special handling of situations in which WISE deviates from its usual
scan pattern to avoid pointing too closely toward the Moon, as described in $\S$5.2 of M17. Our careful handling of such Moon avoidance maneuvers splits what would be overly long coadd epochs into 
pairs of standard coadd epochs occurring immediately before/after each maneuver. WISE scans converge at the ecliptic poles; consequently coverage in these regions is
nearly continuous.

The primary disadvantage of our coaddition is that the parallactic motion within a single coadd epoch's timespan can smear out Planet Nine's appearance and therefore limit the sensitivity of our search. The effect of this limitation on the parameter space we are able to probe is illustrated in Figure 1 of M17, and described fully in $\S$3.2-3.4 of M17. At $|\beta| > 30^{\circ}$, parallactic smearing dictates the minimum $d_9$ for which we claim our search to be sensitive, with this minimum value ramping up from 250 AU at $|\beta| = 30^{\circ}$ to $\sim$600 AU at $|\beta| = 55^{\circ}$. The mean timespan across
all coadd pixels entering our analysis is 1.06 days. The minimum (maximum) number of coadd epochs per \verb|coadd_id| footprint is 7 (12).



\subsection{Coaddition}
\begin{figure*}
        \includegraphics[width=7.0in]{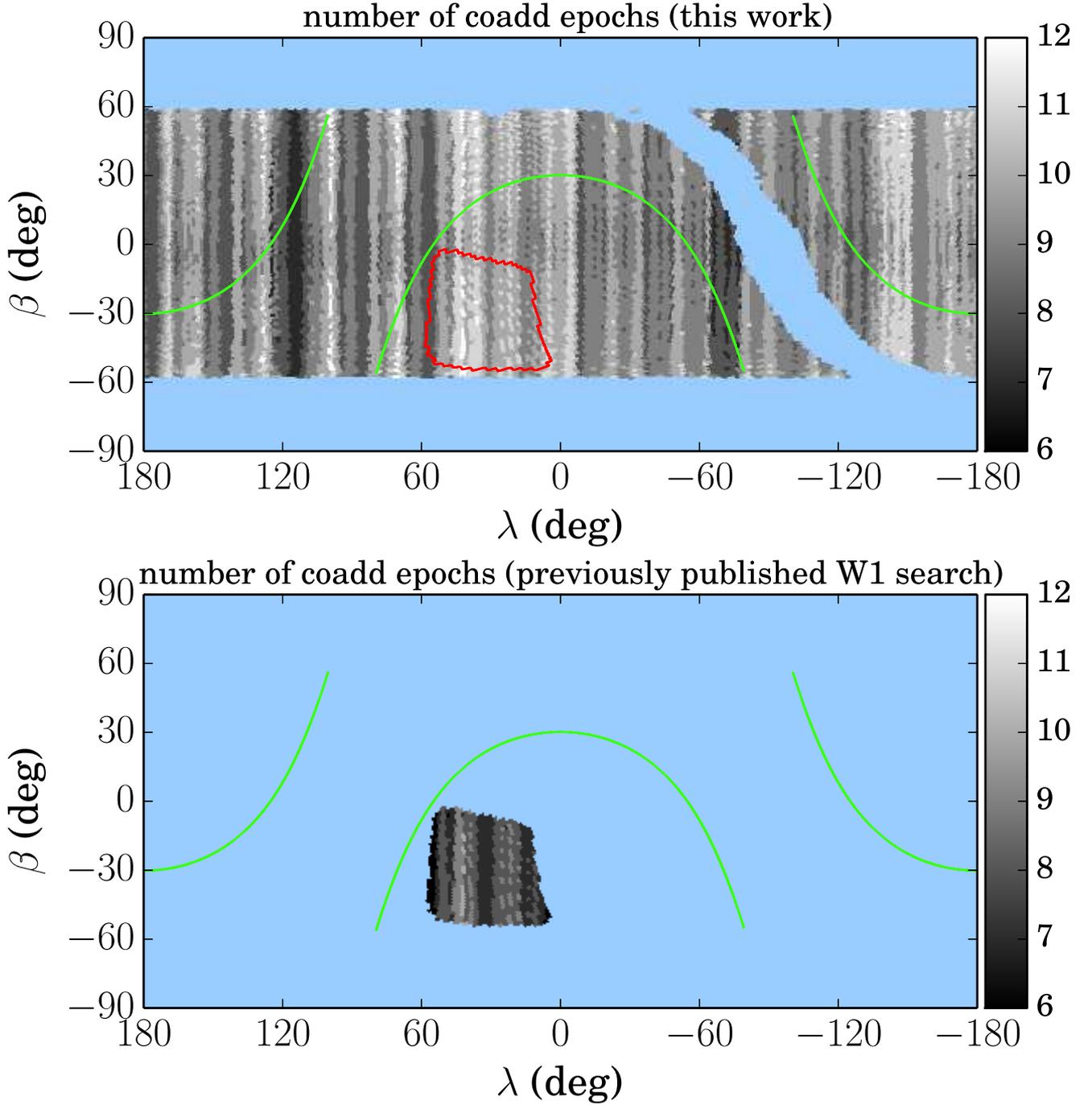}
    \caption{Maps of the number of W1 coadd epochs analyzed, shown in ecliptic coordinates. Light blue indicates sky regions not searched. Top: present search, covering 76\% of the sky (31,480 square degrees) and incorporating $\sim$4 years of W1 observations. Bottom:
    corresponding map for previously published M17 search, for the sake of comparison. M17 searched a much smaller footprint ($\sim$1,840 square degrees), and even
    within that area employed fewer epochs per sky location because the third-year NEOWISER data release was not yet available. The present search incorporates
    129,483 time-resolved coadds, whereas that of M17 incorporated only 6,219 epochal coadds. The entire sky is shown in both panels. As can be seen in the top panel, our
    present search omits tiles at $|\beta| \gtrsim 55^{\circ}$. The diagonal missing stripe centered near ($\lambda$, $\beta$) = ($-$90$^{\circ}$, 0$^{\circ}$) is due to our omission of low $|b_{gal}|$ regions toward the inner Galaxy. The M17 search region boundary is indicated by a red outline in the top panel. Green lines denote $|b_{gal}| = 30^{\circ}$, which we at times adopt as the boundary between `high' and `low' Galactic latitude regions.}
    \label{fig:num_epochs}
\end{figure*}

We employ an adaptation of the unWISE code \citep{lang14} to generate W1 coadds which incorporate the time-slicing described in $\S$\ref{sec:slice} and retain the native WISE angular resolution. Complete details of the unWISE code and the particular options we use in the context of our W1 Planet Nine searches can be found in \cite{lang14}, \cite{meisner16} and M17. The number of coadd epochs generated per sky location is shown in the top panel of Figure \ref{fig:num_epochs}. In total, our present search incorporates 129,483 time-resolved W1 coadds. For comparison, the bottom panel of Figure \ref{fig:num_epochs} shows the number of coadd epochs per sky location in the M17 search, which included only 6,219 such coadds. The median, mode and mean of the number of coadd epochs per sky location are 9, 9, and 8.87 respectively.

Our analysis includes approximately eight full sky passes, plus an additional $\sim$1/6 of a sky pass from early 2011. $\sim$5\% of the sky is missing one of the
usual eight passes due to a WISE spacecraft `command timing anomaly' which halted science observations in 2014 April. As a result, under an idealized survey strategy
in which WISE always points at exactly $90^{\circ}$ solar elongation, we would expect $\sim$80\% of the sky to have 8 coadd epochs, $\sim$15\% to have 9 coadd epochs, and $\sim$5\% to have 7 coadd epochs. However, Moon avoidance maneuvers lead to extra coadd epochs, with each sky location typically affected by one such maneuver.
This results in roughly one `bonus' coadd epoch everywhere on the sky. It is possible that the same sky region has been affected by more than one Moon avoidance maneuver, explaining the Table \ref{tab:n_epoch} tail of $n_{epochs}$ which extends up to 12 over a fractionally small area. The small fraction of area with $n_{epochs} \le 6$ is due to rare
circumstances under which our coaddition fails, for example in the heavily saturated cores of a a few of the brightest stars or during a sighting of Mars, Jupiter or Saturn.

\begin{figure*}
        \includegraphics[width=7.0in]{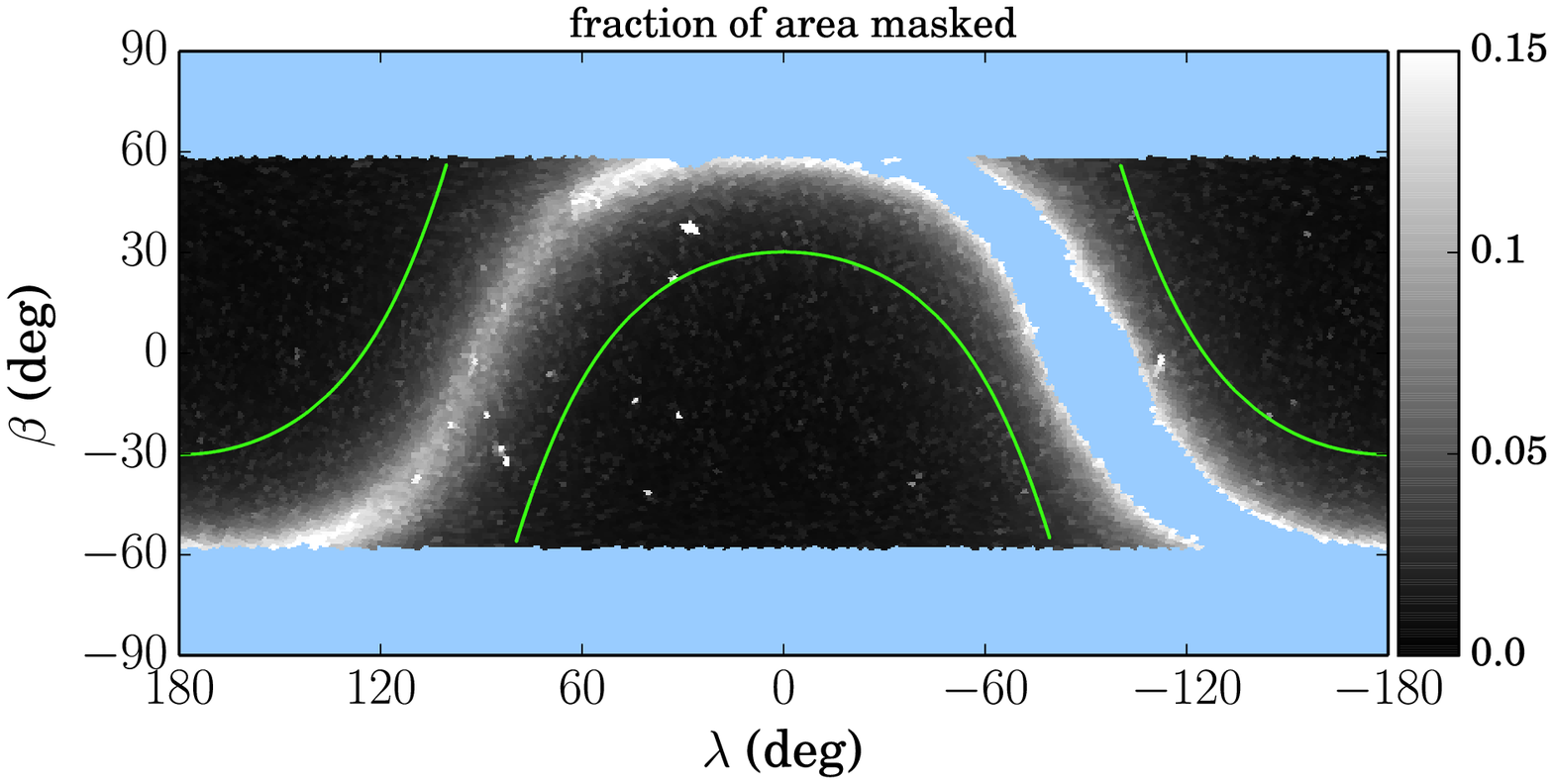}
    \caption{Fraction of area masked as a function of sky location. At high Galactic latitude ($|b_{gal}| > 30^{\circ}$), the typical fraction of area masked is
     $\sim$1\%. This fraction ramps up dramatically toward the Galactic plane, mainly due to larger numbers of 2MASS sources and W1-bright stars
    whose locations are masked by our transient filtering procedure ($\S$\ref{sec:filtering}). Green lines denote the $|b_{gal}| = 30^{\circ}$ boundary.}
    \label{fig:frac_masked}
\end{figure*}

\subsection{Reference Templates}

We create a reference template image corresponding to each time-resolved coadd. The reference template is constructed by combining all coadd epochs of the relevant 
\verb|coadd_id| footprint which are temporally separated by more than 5 months relative to the epoch being searched for transients. Typically, 6--8 epochal coadds contribute to each reference template. This corresponds to a typical reference stack integer coverage of  $\sim$75--100 frames near the ecliptic plane, with this value increasing
toward higher $|\beta|$. The reference templates in this work are slightly deeper than those of M17, but this has only a very minor effect on our search's sensitivity because
the noise in the difference images used for transient detection is dominated by noise in the epochal coadds. Our re-search of the M17 sky region is only substantively better by virtue of incorporating two additional epochs worth of transients, thanks to the third year NEOWISER data now available.


\subsection{Image Differencing}

Our difference imaging pipeline is built around a set of standard astronomical image analysis software packages. Source Extractor \citep{sextractor} and SCAMP \citep{scamp} are used to derive astrometric tweaks that ensure the best possible alignment of the reference and search images. SWARP \citep{terapix} is used to warp the reference based on the SCAMP results. HOTPANTS\footnote{\url{https://github.com/acbecker/hotpants}} is used to perform kernel-matched subtractions that account for small (percent level) changes in the PSF and/or multiplicative differences between the reference stack and the search image.


\subsection{Extracting Transients}

We use Source Extractor to detect transients within our difference images and adopt the \verb|XWIN_IMAGE|, \verb|YWIN_IMAGE| parameters as centroid measurements.


\subsection{Filtering Transients}
\label{sec:filtering}


We go to great lengths to eliminate identifiable artifacts from our transient catalogs while masking the smallest possible fraction of sky area. The transients we remove
include those associated with:

\begin{itemize}
\item static sources, as identified by 2MASS counterparts.
\item kernel matching artifacts. These are ringing features introduced when HOTPANTS attempts to establish consistency between the search and reference image PSFs.
\item the cores and/or diffraction spikes of bright star profiles.
\item latents. These are persistence artifacts which appear at the \textit{detector} position of saturated stars in subsequent exposures. Latents have the appearance of 
diffuse smudges, and are detected as spurious transients because their positions vary as the WISE scan direction changes. Given a catalog of bright star locations it
is possible to very accurately predict the positions of latents. We use the AllWISE catalog to identify bright stars capable of causing latents.
\end{itemize}

In M17, we created custom masks for four `ultra-bright' stars whose profiles extend far beyond the PSF model used in our standard bright star masking approach. In this work, we opted not to continue the practice of making such custom masks for bright stars throughout our search footprint. This choice did not lead to a fractionally significant increase in the number of spurious transients, as the transient catalog is dominated by artifacts associated with much more numerous moderately bright sources, particularly those in relatively low Galactic latitude regions of our search footprint.

At high $|b_{gal}|$ ($|b_{gal}| > 30^{\circ}$), our filtering masks 1.0\% of the area. This fraction varies dramatically over the sky, mainly due to variations
in the number of static background sources within the Galaxy along different sightlines (see Figure \ref{fig:frac_masked}). In $\S$\ref{sec:completeness}, we account for spatial variations of the fractional area masked when assessing our survey's sensitivity.

\subsection{Orbit Linking Overview}
Our linking approach is similar to that of M17, but
different enough in detail that we have fully redescribed it in $\S$\ref{sec:orbfit}. Our orbit linking procedure employs the \verb|orbfit| code of \cite{bernstein00} to test whether groupings of transients (`tuples') can be well-fit by any bound Keplerian orbit in the outer
solar system. This orbit linking methodology is modeled after that of \cite{brown15}, and is capable of linking transients with arbitrary cadence. \textit{Notably, in this work we require that a Planet Nine `candidate' consist of at least five linked transients, versus only four required in M17.} We found this choice to be important in reducing the number of false positive linkages, particularly in crowded regions where the number density of transients spikes dramatically. Because we have incorporated an additional year of
W1 frames relative to M17, we suffer no loss in sensitivity despite requiring a minimum of five rather than four linked transients (see Figure \ref{fig:completeness_high_b}).

The astrometric uncertainties assigned to our transient centroids represent a critical input for orbit linking, as these dictate the $\chi^2$ values returned by \verb|orbfit|. We have
sought to be very conservative/generous in our centroid uncertainty model, which is described in $\S$7.1 of M17. The typical per-coordinate uncertainty we adopt is 0.8$''$, nearly a 
third of a WISE pixel. As in M17, our choice of $\chi^2$ threshold is guided by our desire for a false negative rate of less than 0.001. For quintuplet linkages, this translates to 
a requirement of $\chi^2 < 21.5$. In practice, to be conservative, we have inspected finder charts for all candidates with five or more linked transients and $\chi^2 < 40$. This can
be viewed, in the case of quintuplet linkages, as effectively padding our centroid uncertainties by an additional $\sim$40\%.









\section{Search Region}
\label{sec:footprint}

Although WISE observes the entire sky, some regions have been omitted from our search. The following subsections detail the circumstances under which we have trimmed
area. Our search footprint includes 76.3\% of the sky.

\subsection{High Ecliptic Latitude}
\label{sec:beta}

 As illustrated in Figure 1 of M17, the minimum $d_9$ for which our search method is sensitive increases rapidly at high $|\beta|$. There exists such a minimum $d_9$ value because the timespan of a given W1 coadd epoch increases as 1/cos($\beta$), resulting in excessive parallactic smearing for sufficiently nearby objects at relatively high ecliptic latitude. For instance, the parallactic motion of a body at 400 AU during the course of a single W1 coadd epoch is 1.6$''$ at $|\beta|$ = $10^{\circ}$, but increases to 6.2$''$ by $|\beta|$ = $35^{\circ}$ (see Equation 2 of M17). Excessive parallactic motion within a single coadd's timespan would elongate Planet Nine's appearance, reducing our sensitivity and complicating our completeness simulations. Therefore, our orbit linking ($\S$\ref{sec:orbfit}) is designed to work only for $d_9 > 250$ AU, as parallactic smearing would render our search relatively insensitive at smaller distances, and resolving orbits at smaller distances requires linking transients over progressively wider sky areas, which becomes computationally prohibitive. We note that many postulated Planet Nine orbits have $q < 250$ AU (often $q \sim 200$ AU), with some extending to $q = 150$ AU, though Planet Nine would likely have already been detected if close to perihelion \citep{brown16}. At $|\beta| \approx 30^{\circ}$, the minimum $d_9$ imposed by parallactic smearing becomes larger than 250 AU and therefore sets our search's lower distance limit for $|\beta| > 30^{\circ}$.
 
 The maximum $d_9$ probed is determined by the maximum plausible W1 luminosity of Planet Nine, which corresponds to W1 = 16.1 at a fiducial distance of 622 AU (H$_{W1}$ = 2.13) according to \cite{fortney16}. This translates to a maximum detectable distance that ramps upward from $\sim$800 AU at $|\beta| = 0^{\circ}$ to $\sim$900 AU at $|\beta| \approx 65^{\circ}$.

For $|\beta| < 30^{\circ}$, our survey is sensitive throughout the distance range $250 \lesssim d_9/\textrm{AU} \lesssim 800$. As $|\beta|$ increases beyond 30$^{\circ}$, the distance range within which we are sensitive to Planet Nine narrows rapidly. By $|\beta| \sim 65^{\circ}$ we effectively lose all sensitivity -- the maximum and minimum detectable $d_9$ converge at $|\beta| \sim 65^{\circ}$, assuming H$_{W1} \ge $ 2.13. At $|\beta| \sim 55^{\circ}$, only a relatively narrow range of $d_9$ is probed: $650 \lesssim d_9/\textrm{AU} \lesssim 850$. We therefore restrict our search to tiles with central coordinates interior to the region defined by $|\beta| < 55^{\circ}$, or within 1.1$^{\circ}$ of the $|\beta| = 55^{\circ}$ boundary \footnote{1.1$^{\circ}$ is the distance between the center of a tile and its corners. Therefore, tiles that have centers up to 1.1$^{\circ}$ beyond either $|\beta| = 55^{\circ}$ boundary could potentially contain sub-regions at $|\beta| < 55^{\circ}$.}. This cut removes 3,182 of 18,240 tiles, corresponding to 17.4\% percent of the sky. We note that by inventing 
specially tailored time-slicing rules to define our coadd epoch boundaries at very high ecliptic latitude, we could make use of all WISE data up to $|\beta|$ = 90$^{\circ}$. However, the Planet Nine narrative disfavors very high inclinations \citep{batygin16, batygin17}, and $i > 55^{\circ}$ would be required to observe Planet Nine in the high $|\beta|$ sky area we have excluded.


\subsection{The Galactic Plane}
\label{sec:plane}

We run all stages of our processing up to and inclusive of transient filtering for all 15,058 tile footprints satisfying our ecliptic latitude cut. However, we find that an excessively high number 
density of transients remains in our filtered catalogs in low Galactic latitude regions toward the inner Galaxy. The enhanced transient number density near the Galactic center is due to  the high density of static background sources in this region. Extreme, localized increases in transient density are particularly problematic because of the combinatorics associated with orbit linking -- the number of transient $n$-tuples increases with the $n^{th}$ power of transient number density. We therefore sought to excise regions near the Galactic center containing huge numbers of static background sources. We do so by making a cut on the number of 2MASS \citep{skrutskie06} sources per tile footprint, $n_{2mass}$. We remove tiles with $n_{2mass} > 90$,000. This corresponds to one 2MASS source per 47 pixels. For comparison, the effective number of pixels corresponding to the W1 PSF is 15.8, meaning that our chosen source density threshold translates to roughly 0.34 2MASS sources per WISE beam. This is equivalent to
2.8$\times$10$^{-3}$ 2MASS sources per square arcsecond. We use 2MASS source density as a proxy for crowding, rather than WISE source density,  because 2MASS has far superior angular resolution. In general sources detected by 2MASS are also detected by WISE (WISE is deeper), though deblending issues may modulate/reduce the number of WISE sources actually catalogued in the densest regions.

1,131 tiles are rejected due to our cut on number of 2MASS sources. These tiles trace out a roughly diamond-shaped area centered at ($l_{gal}$, $b_{gal}$) = (0$^{\circ}$, 0$^{\circ}$) when viewed in Galactic coordinates. This region is shown as the empty (light blue) stripe passing through the Galactic center at ($\lambda$, $\beta$) = ($267^{\circ}$, $-6^{\circ}$) in Figures \ref{fig:num_epochs}, \ref{fig:frac_masked}, \ref{fig:completeness_map} and \ref{fig:transient_dens}. The region excised due to high source density reaches a maximum $|b_{gal}|$ of $\sim$$14^{\circ}$ near $l_{gal} = 0^{\circ}$, then tapers toward the Galactic plane, disappearing by $l_{gal}$ = 90$^{\circ}$ toward the east and $l_{gal}$ = 282$^{\circ}$ toward the west. We are able to perform our search all the way into the Galactic plane (through $b_{gal} = 0^{\circ}$) in the outer Galaxy -- our background source density cut is lenient enough to retain most of the outer Galactic plane.



We also found that a region of the Galactic plane near ($l_{gal}$, $b_{gal}$) = (109$^{\circ}$, $-2^{\circ}$) exhibited an unusually high number density of transients
associated with diffuse emission. We presume this to result from time-variable illumination of dust clouds \citep[e.g.][]{rest08}. As a result, we removed eight nearby tiles, with \verb|coadd_id| values 3461p575, 3484p590, 3489p575, 3510p605, 3513p590, 3516p575, 3540p605 and 3542p590.



\subsection{External Galaxies}

External galaxies with exceptionally large apparent sizes also lead to spikes in the transient number density. We therefore excised elliptical regions surrounding M31, Triangulum, M110 and IC10. The combined area of these elliptical external galaxy masks is 6.6 square degrees. The LMC ($\beta \sim -85^{\circ}$) and SMC ($\beta \sim -65^{\circ}$) do not require special masking
because they are already excluded by the ecliptic latitude cut described in $\S$\ref{sec:beta}.


%


After deducting the areas masked at high ecliptic latitude, near the Galactic center, close to time-variable ISM reflection features and surrounding external galaxies, we are left with a 31,480 square degree search region. This corresponds to 76.3\% of the sky. A table summarizing our tile centers and the tile-level cuts described in this section is available online\footnote{\tiny{\url{https://faun.rc.fas.harvard.edu/ameisner/p9w1_3pi/p9w1_3pi-tiles.csv}}}.



\section{Orbit Linking Details}
\label{sec:orbfit}

Our orbit linking strategy follows the procedure we outlined
in M17, with some modification to handle the increased number of
epochs and transients.  First, for each transient, we identify a pool
of all nearby subsequent transients that might be associated with
bound Keplerian orbits at distances of 250 AU or more. Next we construct triplets that consist of the first transient and pairs
of nearby transients from the pool around it. We then test if each triplet
is consistent with a simple model of reflex motion plus linear drift
in the sky plane, and we accept it if its $\chi^2$ is less
than 30. We repeat this process for quadruplets built from accepted
triplets and a fourth transient from the pool. Quadruplets themselves
are tested with the parallax-linear drift model, but with a $\chi^2$
threshold of 150. These steps, which convert the full list of
transients into potential linkages of quadruplets, constitute our
prescreening process.

We test the prescreening process with mock data. By sampling
Keplerian orbital parameters for hypothetical bound solar system
objects beyond 250~AU, we generate mock transients with 
cadence and astrometric uncertainties similar to the real data. Applying the
same parallax and linear drift tests on triplets and quadruplets as we
do for real data, we find that our prescreen step falsely rejects no
more than 1 in $10^4$ linkages. The lost linkages are simply cases in which the simulated statistical errors 
happen to perturb the simulated Keplerian trajectories in such a way as to end up with poor $\chi^2$. This
fractional loss is small enough to be consistent with our overall goal of keeping the false negative rate due to
statistical errors on transient centroids below 0.001.

In the next step of the linkage procedure, we evaluate each
prescreened quadruplet with \verb|orbfit|, a code designed
by \cite{bernstein00} to fit observations of outer solar
system objects with Keplerian orbital elements. We choose an \verb|orbfit| $\chi^2$
threshold of 40 in the decision to accept or reject a linkage,
irrespective of tuple order or degrees of freedom (\verb|orbfit| uses several
different models, depending on the details of the transients). This
choice yields false negative rates that are well below the desired maximum value of 0.001
adopted in M17. We began the present search using $\chi^2$ thresholds yielding false negative
rates of exactly 0.001, to match M17, but expanded our analysis to higher $\chi^2$ values in attempt to be
conservative. The actual value of 40 is not special, and was chosen simply because it corresponds to a false negative
rate below 0.001 while yielding an ambitious yet tractable number of candidates requiring finder chart inspection.

Finally, we check if the quadruplets that are admitted on the basis
of \verb|orbfit| have any transients in common. In cases where they do, we
merge them together to form quintuplets, or even
higher-order \'tuples\', if the merged \verb|orbfit| $\chi^2$ is less than
40, as before. For all of the linkages that we find, the number of
degrees of freedom is low enough that this $\chi^2$ threshold
preserves a false negative rate below the 0.001 threshold.

This procedure culls over 5 million transients into roughly 2,500 quintuplets, many of which are subsets of nearly 200 higher-order $n$-tuples ($n$ = 6$-$8; see $\S$\ref{sec:results}). Along the way, the prescreen process enumerates over 92 billion triplets and 275 million quadruplets. Our code ends up calling \verb|orbfit| to check over 8 million potential linkages.




\section{Completeness Analysis}
\label{sec:completeness}

Our completeness estimates are based on an empirical detection efficiency measured from 300,000 fake sources we have injected into our real difference images. We measure the fraction of these fake sources recovered as a function of W1 magnitude and number of exposures ($n_{exp}$) contributing to the relevant epochal coadd pixels. Using the probability
distribution functions for both $n_{exp}$ and the number of coadd epochs per sky location ($n_{epoch}$), we then perform Monte Carlo simulations to assess the probability with which five or more transients will be recovered as a function of an object's W1 magnitude. This latter probability is what we refer to as our completeness, since our search demands that a Planet Nine candidate consist of a low-$\chi^2$ linkage containing at least five transients. Our Monte Carlo simulations also account for the fraction of area masked ($f_{masked}$) by randomly zeroing out the appropriate (spatially varying) fraction of single-epoch detection probabilities.

\begin{figure}
        \includegraphics[width=3.3in]{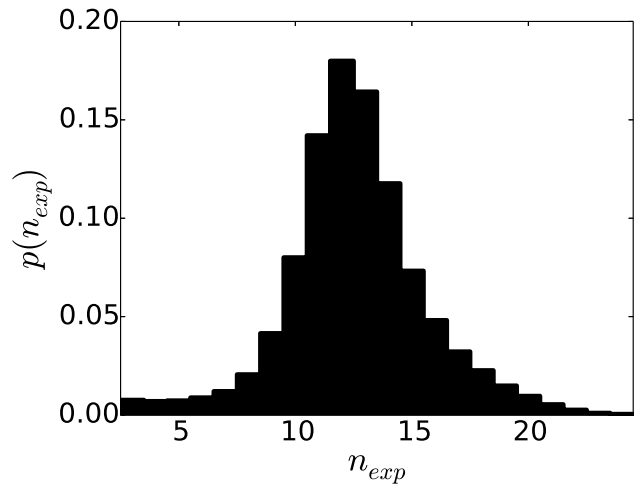}
    \caption{Probability distribution function for the number of exposures per coadd pixel, for coadd pixels entering into our analysis.}
    \label{fig:n_exp}
\end{figure}

The probability distribution function for $n_{epoch}$ is provided in Table \ref{tab:n_epoch}. We have also computed the probability distribution function of $n_{exp}$ for the 129,483 tiles in our present search, and show the results in Figure \ref{fig:n_exp}. Both the median and mode are $n_{exp} = 12$.

\begin{figure}
        \includegraphics[width=3.3in]{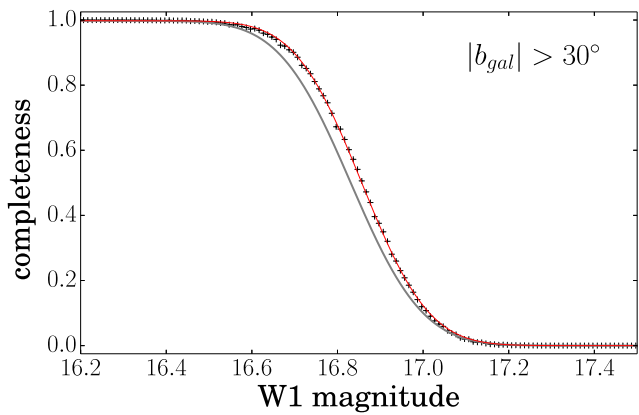}
    \caption{Average completeness curve for our search, restricted to high Galactic latitude. The plus marks show the results of our completeness simulations, and the red line is
    a best-fit error function model (see Equation 10 of M17 for the parametrization that we use). The grey line is the best-fit model from the M17 search. The completeness of this search and the M17 search are similar because
    we have folded in an additional year of W1 data relative to M17, but also require one additional detection per linkage relative to M17 (in this work we require quintuplets whereas M17 required quadruplets). The
    average high-latitude completeness curve reaches 90\% (50\%) completeness at W1 = 16.70 (16.86).}
    \label{fig:completeness_high_b}
\end{figure}

Figure \ref{fig:completeness_high_b} shows the results of our Monte Carlo simulations at high Galactic latitude ($f_{masked} = 0.01$), plotting the fraction of objects with $\ge$5 transient detections as a function of W1 magnitude. 90\% (50\%) completeness is achieved at W1 = 16.70 (16.86). $\S$8 of M17 provides additional details of our fake source injection ($\S$8.3) and Monte Carlo simulation procedures ($\S$8.5).


Because our search spans a wide range of background source densities and includes sub-regions with different $n_{epoch}$ values, we have also produced a map of the 90\% completeness threshold's spatial variation. In doing so we perform additional Monte Carlo simulations to simultaneously account for five factors:

\begin{enumerate}
\item The local value of $n_{epoch}$ as provided by the map shown in Figure \ref{fig:num_epochs}. Regions with more coadd epochs will be more sensitive and therefore reach 90\% completeness at higher W1 magnitudes.
\item The local value of $f_{masked}$ as given by the map shown in Figure \ref{fig:frac_masked}. Regions with larger $f_{masked}$ have reduced sensitivity.
\item Additional background noise in regions near the Galactic plane, presumably due to large numbers of static sources, both resolved and unresolved, as well as diffuse Galactic
emission. We find that this effect is well-approximated by a reduction in depth of:

\begin{equation}
\begin{split}
\Delta_{bg} = -1.08\times10^{-2} + (2.76\times10^{-6})n_{2mass} + \\ 
(5.94\times10^{-12})n_{2mass}^2 \ \textrm{[mag]}
\end{split}
\end{equation}

This effect is largest in the most crowded regions of our footprint, reaching a maximum value of 0.29 magnitudes.


\item Variation in the amount of parallactic smearing as a function of ecliptic latitude, resulting from the fact that the timespan of a single coadd epoch scales as 1/cos($\beta$). This tends to reduce the sensitivity of our search at high $|\beta|$ relative to low $|\beta|$.
\item The average increase of $n_{exp}$ as 1/cos($\beta$). This effect tends to increase the sensitivity at high ecliptic latitude relative to that at low ecliptic latitude.
\end{enumerate}

Figure \ref{fig:completeness_map} shows the variation in 90\% completeness threshold implied by taking these effects into account, for a fiducial distance of $d_9$ = 700 AU. The map
is a function of $d_9$ because of the parallax-dependent term described in item (iv) above. The vertical stripes in Figure \ref{fig:completeness_map} trace variations
in $n_{epoch}$ as a function of ecliptic longitude. Reduced sensitivity toward the Galactic plane is also apparent, due to the effects described in items (ii) and (iii) above. The
ecliptic latitude dependent effects in items (iv) and (v) tend to cancel each other out and are not especially obvious in the Figure \ref{fig:completeness_map} map. The mean 90\%
completeness threshold within our search footprint is W1 = 16.69, and the 25$^{th}$ (75$^{th}$) percentile value is W1 = 16.64 (W1 = 16.76). The full range of 90\% completeness threshold values spanned is quite broad, 15.97 $ < \textrm{W1} < $ 16.93. The map shown in Figure \ref{fig:completeness_map} is available
online\footnote{\tiny{\url{https://faun.rc.fas.harvard.edu/ameisner/p9w1_3pi/completeness_map_fullsky-0128.fits.gz}}} in FITS format.






\subsection{Cautionary Notes on Completeness}

\begin{figure*}
        \includegraphics[width=7.0in]{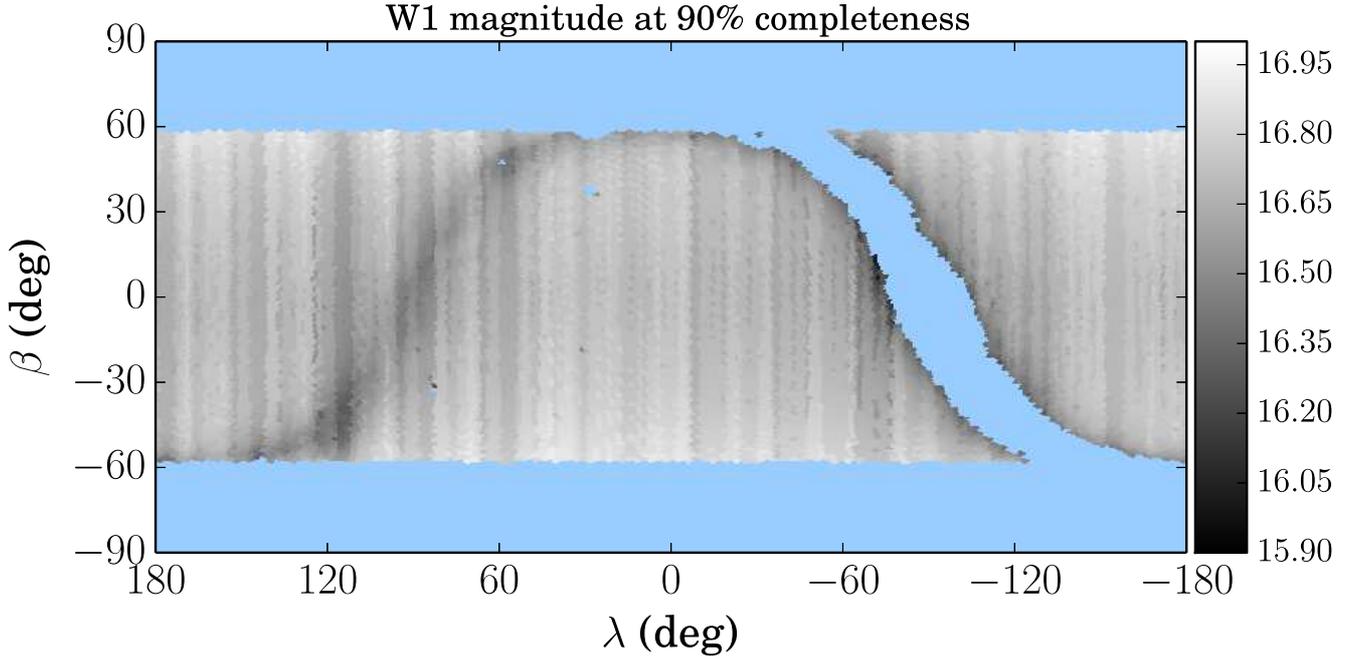}
    \caption{Map of our search's 90\% completeness threshold, for a fiducial present-day Planet Nine distance of 700 AU. The vertical stripes are due to variations in the
    available number of coadd epochs with ecliptic longitude (see Figure \ref{fig:num_epochs}). Our reduced sensitivity near the Galactic plane is also apparent.}
    \label{fig:completeness_map}
\end{figure*}

As described in $\S$8.5 of M17, saturation and/or our bright source masking could cause us to miss Planet Nine in the highly unlikely event that it has $W1 < 9.5$. Additionally, 
our Monte Carlo analysis does not capture spatial correlations between the locations of masked pixels -- it is certainly possible that a body at many hundreds of AU could remain
hidden in front of e.g. a single extremely bright star during the entire timespan of the WISE mission. The typical fraction of area masked at high Galactic latitude is $\sim$1\%. Our orbit linking can pair transients up to $\sim$1.35$^{\circ}$ apart in order
to ensure recovery of objects as close as 250 AU, so that within $\sim$1.35$^{\circ}$ of our search region's boundaries (e.g. near $|\beta|$ = 55$^{\circ}$) there are additional caveats
to our 90\% completeness values shown in Figure \ref{fig:completeness_map}. The largest value of $d_9$ for which we claim our completeness analysis applies is dictated by self-subtraction. Our conservative upper limit is $d_9 = $ 2,250 AU, assuming $e_9 \le 0.75$, as explained in $\S$3.6 of M17.




\begin{figure}
        \includegraphics[width=3.3in]{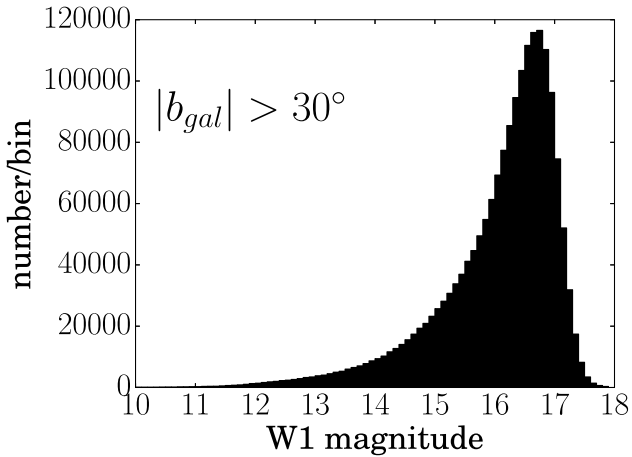}
    \caption{Histogram of W1 magnitudes for detections in
our filtered transient catalog. The source counts peak near W1=16.6, far fainter than the single-exposure detection limit at W1 = 15.3.}
    \label{fig:transient_mags}
\end{figure}


\section{Results}
\label{sec:results}

\begin{figure*}
        \includegraphics[width=7.0in]{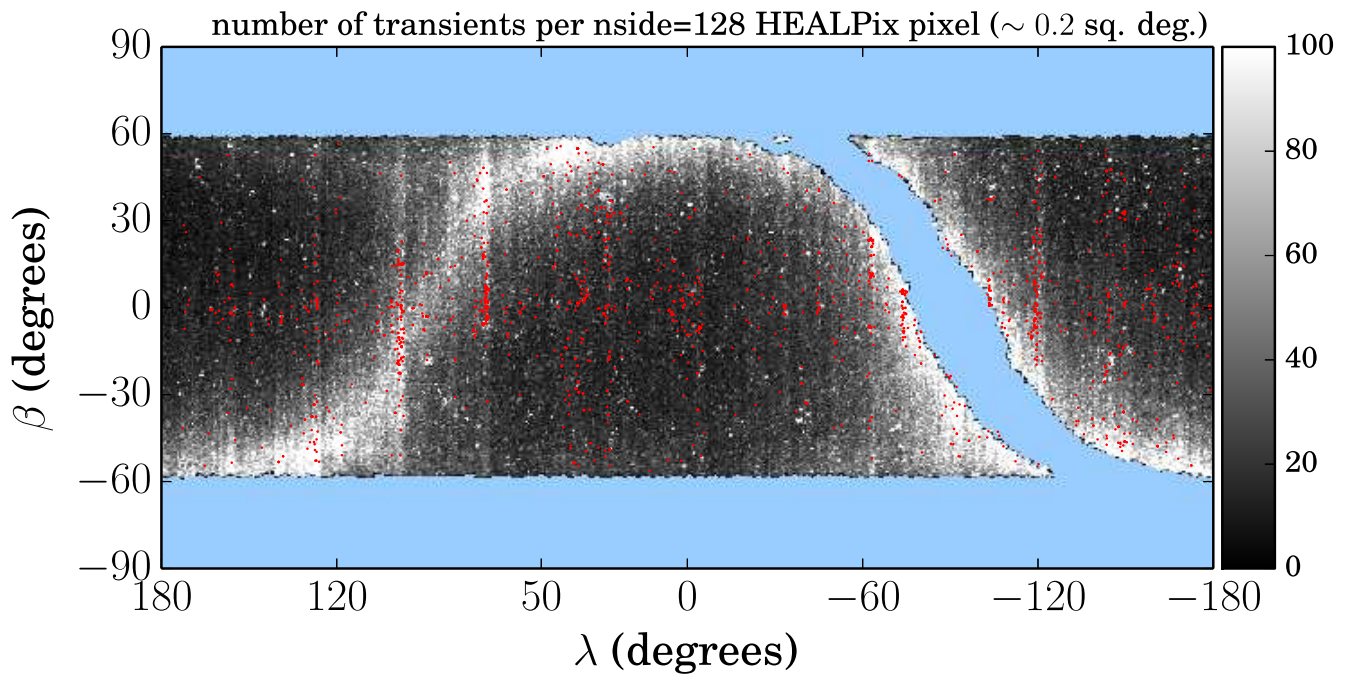}
    \caption{Number density of transients throughout our search footprint. The scale bar is in units of transients per $N_{side}$ = 128 HEALPix pixel. Each such pixel
    is $\sim$0.2 square degrees in size. The locations of our 2,560 quintuplet linkages with \texttt{orbfit} $\chi^2 < 40$ are indicated by red dots.}
    \label{fig:transient_dens}
\end{figure*}

5,603,879 transients remain after filtering our difference image detections as described in $\S$\ref{sec:filtering}. Figure \ref{fig:transient_mags} shows the distribution of W1 magnitudes for our transients at high Galactic latitude. As expected, this distribution peaks near W1 = 16.6, which is $\sim$1.3 magnitudes below the
single-exposure detection limit at W1 = 15.3. The spatial distribution of these transients is shown in Figure \ref{fig:transient_dens}. 


Running our orbit linker on the full set of transients results in 2,735 linkages with \verb|orbfit| $\chi^2 < 40$ and a minimum of five transients. 2,560 of these linkages are quintuplets, 163 are sextuplets, 11 are septuplets, and one is an octuplet. The locations of all 2,560 low-$\chi^2$ quintuplet linkages are indicated by red dots in Figure \ref{fig:transient_dens}. Because WISE scans are oriented vertically in this coordinate system, essentially all relevant quantities (such as the number of available coadd epochs) and systematics (such as periods of relatively high Moon glow) will imprint as vertical stripes. The transient density and therefore the density of false positive quintuplets also trace these vertical stripes. For instance, the vertical stripes in Figure \ref{fig:num_epochs} will cause vertical striping in the number of transients per unit area, simply because certain stripes have more or fewer coadd epochs available.

We have examined finder charts for all 2,735 ($\chi^2 < 40$) candidates. We note that this $\chi^2$ threshold is quite conservative in terms of its corresponding false negative rate. For instance, had we implemented cuts to achieve a formal false negative rate of 0.001, only 1,296 linkages would require finder chart inspection.

In scrutinizing the finder charts, we seek to determine whether any linkage exists for which $\ge$5 of its transients are legitimate, in the sense that they cannot be ruled out as associated with artifacts or previously known moving objects. Because we do not impose any cuts on the orbital elements returned by \verb|orbfit|, it turns out that many of our linkages are known high proper motion objects in the solar neighborhood. We therefore downloaded the list of all SIMBAD \citep{simbad} proper motion sources with $|\mu| > 200$ mas/yr, and used this catalog
to automatically flag linkages associated with these known moving objects, employing a match radius of 15$''$. 737 of our 2,735 linkages are associated with known high proper motion objects cataloged by SIMBAD. We exercised caution by checking each such finder chart to ensure a legitimate association, and verified that the direction of proper motion seen in our finder chart matched that quoted by SIMBAD. A small number of our linkages displayed linear motion consistent with being stars in the solar neighborhood, and appeared to be new discoveries upon searching SIMBAD and Vizier \citep{vizier}. In these cases we checked 2MASS data to verify the linear nature of the motion on longer timescales.

For all remaining 1,998 linkages, we were able to rule out a sufficient number of transients such that four or fewer legitimate transients remained. The spurious transients that we ruled out can be grouped into three dominant categories:

\begin{itemize}
\item Latents. It is difficult to automatically flag 100\% of latents for several reasons. For instance, we found a number of cases in which the parent bright star is entirely absent from the AllWISE catalog from which we constructed a sample of potential latent parents. In a few similar cases, the parent bright star is present in AllWISE yet has a W1 magnitude listed as `null'. In many cases the parent's AllWISE W1 magnitude is slightly fainter than
the brightness threshold we employed in constructing our sample of latent parent stars. Latents of third order or higher often contaminated our linkages. Such latents occur $\ge$3 exposures after imaging of the parent star, whereas our automated masking of latents only accounts for first and second order latents.
\item Diffraction spikes. Our bright star masks are highly detailed and do account for diffraction spikes, but these masks are based on a W1 PSF model of angular size 14.9$'$. Even 
though this PSF model extends quite far into the wings, the diffraction spikes of exceptionally bright stars can remain detectable far beyond the angular extent of our masks.
\item Known high proper motion objects which evaded our SIMBAD crossmatch. This includes cases where a known proper motion counterpart was found in Vizier but not SIMBAD and
extremely fast-moving objects ($|\mu| \gtrsim 1.5''/\textrm{yr}$) for which our 15$''$ SIMBAD match radius was insufficient.
\end{itemize}

Other relatively common contaminants included spurious transients associated with bright star glints, real astrophysical variability of static background sources and kernel matching
artifacts which evaded our filtering procedure.

One might view the higher-order linkages (sextuplets and higher) as the most interesting candidates, since in these cases the number of linked transients exceeds our minimum value of five. However, we found that all sextuplets, septuplets and the one octuplet were especially clearcut cases in which sufficient numbers of transients could easily be ruled out as
known high proper motion objects or bright star artifacts. In particular, the octuplet consisted of eight detections of the white dwarf 2MASS J14205484$-$0905086.

As we do not find any valid linkages of $\ge$5 transients, we are left with a non-detection of Planet Nine and the completeness constraints described in $\S$\ref{sec:completeness}.




\section{Conclusion}
\label{sec:conclusion}

We have performed the deepest and widest area WISE-based search for Planet Nine, covering over three quarters of the sky. At high Galactic latitude, we rule out the presence of Planet Nine in the parameter space searched with $W1 < 16.7$ (16.86) at 90\% (50\%) completeness. The sensitivity of our survey is reduced toward the Galactic plane, as illustrated in Figure \ref{fig:completeness_map}. Our search methodology is valid for $250 < d_9/\textrm{AU} < $ 2250, although the lower distance boundary is
dependent on ecliptic latitude. For the most W1-luminous \cite{fortney16} model (which has a mass of 10M$_{\oplus}$), our search probes distances up to 800$-$900 AU depending on ecliptic latitude. The optical-W1 color of Planet Nine is extremely model-dependent, and is further complicated by the fact that intrinsic emission in W1 becomes
fainter as $d_9^{-2}$ whereas reflected sunlight in the optical becomes fainter as $d_9^{-4}$. For the maximally W1-luminous \cite{fortney16} model, our $W1$ fainter than 16.7 constraint corresponds to a limit of $\textrm{VR}$ fainter than 22.1 at a fiducial distance of 650 AU.

Our methodology should be readily applicable to the WISE W2 channel. Typically, (W1$-$W2) $\approx$ 0.8 corresponds to equal signal-to-noise in W1 and W2, meaning that we should expect sensitivity to Planet Nine at W2 $< 15.9$.  This survey limit would render the most W2-luminous model of \cite{fortney16} detectable at distances of up to $\sim$1,700 AU. One operational advantage of searching in W2 rather than W1 is that W2 has a factor of $\sim$2 fewer background sources relative to W1. A disadvantage is that scattered moonlight,
one of the dominant time-dependent artifacts affecting WISE data, is more pronounced in W2 than in W1.


Whereas our survey has omitted some regions, we note that the ongoing \cite{kuchner17} search is examining the entire sky in both W1 and W2, based on the time-resolved coadds of \cite{tr_neo2}. Because the \cite{kuchner17} search employs deep W2 coadds, it can also serve as a fainter extension of the \cite{luhman14} search for a roughly Jovian mass companion to the Sun. WISE continues to collect additional data in W1 and W2, so that there will eventually be at least one additional year of NEOWISER exposures in both bands relative to those currently available.


\section*{Acknowledgements}

This work has been supported by grant NNH17AE75I from the NASA Astrophysics Data Analysis Program, the {\it NASA Outer Planets Program} through grant NNX11AM37G and {\it Emerging Worlds} program grant NNX17AE24G. Resources supporting this work were provided by the NASA High-End Computing (HEC) Program through the NASA Center for Climate Simulation (NCCS) at Goddard Space Flight Center. We thank Eddie Schlafly for helpful comments on this manuscript.

This research makes use of data products from the Wide-field Infrared Survey Explorer, which is a joint project of the University of California, Los Angeles, and the Jet Propulsion Laboratory/California Institute of Technology, funded by the National Aeronautics and Space Administration. This research also makes use of data products from NEOWISE, which is a project of the Jet Propulsion Laboratory/California Institute of Technology, funded by the Planetary Science Division of the National Aeronautics and Space Administration. This research has made use of the NASA/ IPAC Infrared Science Archive, which is operated by the Jet Propulsion Laboratory, California Institute of Technology, under contract with the National Aeronautics and Space Administration.

The National Energy Research Scientific Computing Center, which is supported by the Office of Science of the U.S. Department of Energy under Contract No. DE-AC02-05CH11231, provided staff, computational resources, and data storage for this project.

\bibliographystyle{mnras}
\bibliography{p9w1_fullsky}

\end{document}